# Smart Contracts that are Smart and can function as Legal Contracts

# A Review of Semantic Blockchain and Distributed Ledger Technologies

**Marcelle von Wendland, Bancstreet Capital Partners Ltd**

mvw@alumni.lse.ac.uk or mvw@bancstreet.com

Blockchain and Distributed ledger Technologies are increasingly becoming key enablers for vital innovation in financial services, manufacturing, government and other industries. One of the biggest challenges though is the level of support for semantics by most of the Block Chain and Distributed Ledger technologies. This paper reviews and categorises common block chain and DLT approaches and introduces a new approach to Blockchain / DLT promising to resolve the semantic problems inherent in other Blockchain / DLT approaches.

## 1. Introduction

*Semantic Blockchain as a name combines two of the hottest buzzwords of the 2010's. To the casual observer this may look like hype breeding hype. A deeper examination however reveals a very different picture. While the underlying topic areas "Semantic" and "Blockchain" have been evolving over centuries and decades respectively the meteoric rise of their importance as fields of study and practical application is neither accidental nor coincidental. Both are different aspects of the monumental transformation our global society is undergoing as we move more and more social, political, legal, economic and technical interactions and transactions into new virtual, dematerialised forms underpinned by the capabilities of digital technology. Almost all such interactions and transactions require the ability for participants to obtain two types of certainty: First is the certainty that the meaning of key communications is the same for all participants at critical points during an interaction and that all critical elements of a transaction have the same meaning to all participants. The second certainty required is that there is certainty about whether and under what circumstances agreement has taken place between participants in an interaction or transaction. It is worth to consider each in turn.*

### 1.1 The Need for Certainty of Meaning

The requirement for certainty of meaning is so intuitive and so fundamental that it is often taken for granted. Every type of social, political, legal, economic and technical interaction or transaction has informal and/or formal protocols for achieving certainty of meaning at critical points. Often participants are not even fully aware of these protocols or how they work but this does in no way diminish their critical importance. As we create digital twins of existing interactions or transactions or even create entirely new digital interactions or transactions we need to re-engineer these protocols or create them from scratch. The term **semantics** as widely understood today refers to this process of creating of such digital protocols for getting certainty of meaning. The seminal **[Berners-Lee et al 2001]** article on the semantic web not only signposted the rise of activity in this field also highlighted the fact that digital networks and the digital interactions and transaction they enable can and must be supported by digital means for establishing certainty of meaning. Since then a new cottage industry has arisen around the creation of digital ontologies and the theoretical insight, methods, notations and tools needed for their construction. The present book is just another sign of this.

It is worth considering though whether certainty of meaning by itself is enough and would also mean participants have certainty of agreement. An indicator that it may not be the case is that the combination of semantics with blockchain is more recent and research activity started to increase in the early 2010's. **[Ugarte 2017]** provides a great account of some of this early research as well as details on how semantic web concepts like linked data and digital ontologies based on OWL can and have been applied to financial and other commercial interactions and transactions in combination with block chain technologies such as Bitcoin and Ethereum. Indeed referring to **[Berners-Lee et al 2006]** Ugarte **[Ugarte 2017, p1]** points out that from the 2005 onwards there was a realisation that semantics alone was not the answer. In the article referred to, Tim Berners Lee sets out his vision:

*".I have a dream for the Web [in which computers] become capable of analysing all the data on the Web: the content, links, and transactions between people and computers. A 'Semantic Web', which should make this possible, has yet to emerge, but when it does, the day-to-day mechanisms of trade, bureaucracy and our daily lives will be handled by machines talking to machines. The 'intelligent agents' people have touted for ages will finally materialize..."*

Semantics and digital ontologies are the way to allow machines to obtain certainty of meaning in interactions and transactions. Both Ugarte and Tim Berners Lee make it clear that as we move to a digital world certainty of meaning in a digital context also requires in addition certainty of agreement in a digital context. Before we examine semantics and how it helps machines determine certainty of meaning we therefore need to examine the need for certainty of agreement

## 1.1 The Need for Certainty of Agreement

At an intuitive level it is clear that two participants to a disputed transaction may very well have agreed on the precise meaning of every aspect of the transaction. It could even be that both parties have clear digital evidence that the precise meaning of each and every aspect of the transaction was shared by both of them. But even in that case further evidence is required to ascertain that they both intended and in fact did enter into an agreement on the transaction in question. In other words, agreement is a process, separate and distinct from the meaning of a transaction or the meaning of individual communications in an interaction. A transaction including any form of contract only becomes significant if certain protocols are followed in the interaction between the parties concerned. Certainty of meaning w.r.t to each of the communications relevant to such a protocol is only a necessary precondition but not sufficient in itself for proving that the protocol establishing agreement in the context was indeed adhered to. Creating a digital twin of the protocols that are used in existing human interactions however, is not a trivial challenge Computers have for decades been used to record, transmit or in some form process contracts and other agreements concluded by people. Under certain highly controlled circumstances and with suitable systems and arrangements computers have more recently also started to be used for forming agreement on behalf of the trading partners in situations such as electronic trading between highly trusted partners. Unfortunately the approaches used in those cases do not solve the more general problem that is at the heart of Tim Berners-Lee's vision above and central to fully digital systems not only in Capital Markets, Banking and Financial Services but also in Government, Commercial and Industrial Supply chains, many IOT (Internet of Things) applications and beyond.  The obstacle that needs to be overcome is that in the general case there are no carefully constructed and maintained arrangements in place between two or more parties that want to form an agreement on a transaction either adhoc or as part of a more complex longer running interaction. The problem to be solved is also known as the "Byzantine Generals Problem" and is well described **[Lamport et al 1982],** a paper with the same name.  The Byzantine Generals Problem describes the situation of participants who want to have a trusted conversation between each other to reach a consensus decision but are isolated from each other and can only communicate with each other via messages using channels that by themselves are not trustworthy. Leslie Lamport also presented a solution to the problem in **[Lamport et al  1998]**  that together with the solution presented in **[Liskov et al 1999] and**  developed independently by Barbara Liskov and colleagues has shaped much of the subsequent research. Early papers like **[Cachin 2001]**  were quick to point out applications and such work prepared the ground for HyperLedger, one of the alternatives in the BlockChain space.  The other two alternatives that are available at this time are **proof-of-work (PoW)**  and **proof-of-state (PoS) algorithms.**  *A nice summary of the three approaches is available in [Hammerschmidt 2017].*  One of the early application **PoW** as a consensus mechanism was in HashCash, described by Adam, Back in **[A.Back 2002]**. Satoshi Nakaomoto's  Bitcoin as described in **[Nakamoto 2009]**   then built on this earlier work and also uses PoW.  Proof-of-state (PoS) algorithms were developed later to address some of the draw backs of PoW and particularly its inefficient use of energy as described in **[Laurie 2011] .**  One of the early adopters of was PPCoin described in **[King et al 2012]** and further work by Vitalik Buterin **[Buterin2014]**   and Gavin Wood **[Wood 2014]**  lead to Ethereum which also moved beyond providing a mere crypto coin and provided its own mechanism for creating Smart Contracts directly as part of Ethereum itself. All three approaches, PBFT, PoW and POS continue to be used in the Blockchain space but PBFT and POS are of most interest because of their much higher efficiency.

## 1.3 Combining Blockchain and Semantics

Having looked at the need for both certainty of meaning and certainty of agreement and some of the general solutions for each it is now worth considering how blockchain and semantics can be combined in practice. There are two general ways: First it is possible to create a blockchain mechanism that allows smart contracts or other protocols to be defined using a way that mimics a Turing Machine like eg a microprocessor; the instructions here are telling the mechanism exactly HOW to compute a result but provide no direct insight into what is required. This could be called *semantic blockchain with procedural semantics.*  The second approach is to create a block chain mechanism that takes instructions in the form specifications of the required results but without specifying exactly how the result is to be computed; The instructions here specify exactly WHAT is required but leave it to the mechanism to find the precise way for HOW to compute the required result. This could be called *Semantic blockchain with declarative semantics.*  It is worth to first consider semantic blockchain with procedural semantics in the next section because it is now widely used in approaches like Ethereum and HyperLegder and then explore how *Semantic blockchain with declarative semantics works and solves some of the challenges arising the context of procedural semantics.*

## 2. Semantic Blockchain with Procedural Semantics

Early Blockchain efforts were either focussed on digital cash like Bitcoin **[Nakamoto 2009]**, controlling resource use like HashCash **[A.Back 2002]** or as in **[Lamport et al 1998]**, and **[Liskov et al 1999]**, Byzantine Fault Tolerant state machine replication computing primitives to be engineered into wider solutions. Semantics in those early effort was either fixed and implied as in Bitcoin and hash cash or assumed external to the mechanism as in Lamports PAXOS and Liskovs PBFT.

In the early 2010's researchers and practitioners realised that the computational semantics of platforms like Bitcoin could be used to construct a wide variety of applications. **Hal Hodson's article** "Bitcoin moves beyond mere money" **HODSON 2013]** in the New Scientist provide an early overview of this activity. However while Bitcoin allows a certain amount of scripting directkly as part of the architecture more complex smart contract require mechanisms to be grafted onto Bitcoin. This realisation lead researchers and practitioners to explore ways in which a broader scripting language could be embedded into new coin designs. In **[Buterin 2014],** Vitalik Buterin describes how Ethereum had been specifically designed for this purpose. **[Bartoletti et al 2017]** provides a broad survey of computational semantics embedded into coins like Bitcoin and Ethereum and their use for constructing smart contracts.

At the same time, also starting in the early 2010, researchers and practitioners also started to look for alternative ways to implement block chain style smart contracts without using coins. Following a line ealier set out by **[Cachin 2001]** one of the best know project that took this direction is HyperLegder. It was created by Dan O'Prey and Daniel Feichtinger (see [Swanson 2016]) and uses a Practical Byzantine Fault tolerance (PBFT) approach ( see [Liskov et al 1999] ) to provide a distributed legder that can be used either simply as a ledger or augmented with a procedural mechanism called *chain code* **[Cachin 2017]** to realise smart contracts.

Both the coin based approaches mentioned above and approaches like HyperLegder employ procedural or imperative semantics when it comes to implementing actions like smart contracts. Accoerding to "**imperative programming** is a [programming paradigm](#) that uses [statements](#) that change a program's [state](#)" **[WIKIPEDIA01]** This means that any action in a smart contract is defined in a language similar to either machine code (assembler) or higher-level languages like or similar to C/C++/Java etc. In an imperative program the meaning or intended effect of any action in terms of an application domain is implicit and if an explicit form of the meaning is required it must be synthesised. While imperative blockchain scripts (e.g. Ethereum Code **[Wood 2017]**, HyperLegder Chain Code **[Cachin 2017 ]** ) can and indeed invariably do have carefully constructed computational semantics – that is an ontology of the different code constructs the scripting language provides such an ontology does not allow for direct representation of the business meaning of any specific script.

Imperative blockchain scripts can of course still use carefully constructed domain ontologies for representing data but excluding action semantics means that the burden for ensuring the correct action semantics is firmly assigned to the designer and programmer of a particular smart contract.

The participants to such a smart contract must rely on the representations of the designer, programmer or knowledgeable evaluator when it comes to result or business meaning of each imperative action script in a smart contract using an imperative blockchain script. It is now worth to consider the action semantics first of Ethereum and the HyperLegder in more detail.

## 2.1 A Semantic Blockchain Turing Machine - The Ethereum Approach

The Ethereum yellow paper **[Wood 2017]** carefully defines the Ethereum virtual machine (EVM) and its language (opcodes). Being a stack machine the EVM presents a language has vocabulary and semantics that is essentially the same as that of microprocessors. Given its specialist purpose and the fact that it is a virtual stack machine rather than a physical microprocessor its instruction set is more compact and has a few specialist instructions. Its arithmetic, comparison, bitwise logic and most of its stack, memory and control flow operations are exactly what you would expect to find in a micro processor. (see Figure 1 & 2 below) In addition the EVM introduces specialist hash and block chain operations as well as instructions to ensure e.g. that a contract can only be executed once.

The specialist block chain operations like eg CREATE for creating an account, CALL, for sending a message to an account and CALLDATALOAD for loading data from the environment allow for

0s: Stop and Arithmetic Operations

0x00    STOP        Halts execution
0x01    ADD         Addition operation
0x02    MUL         Multiplication operation
0x03    SUB         Subtraction operation
0x04    DIV         Integer division operation
0x05    SDIV        Signed integer
0x06    MOD         Modulo
0x07    SMOD        Signed modulo
0x08    ADDMOD      Modulo
0x09    MULMOD      Modulo
0x0a    EXP         Exponential operation
0x0b    SIGNEXTEND  Extend length of two's complement signed integer

10s: Comparison & Bitwise Logic Operations

0x10    LT       Lesser-than comparison
0x11    GT       Greater-than comparison
0x12    SLT      Signed less-than comparison
0x13    SGT      Signed greater-than comparison
0x14    EQ       Equality comparison
0x15    ISZERO   Simple not operator
0x16    AND      Bitwise AND operation
0x17    OR       Bitwise OR operation
0x18    XOR      Bitwise XOR operation
0x19    NOT      Bitwise NOT operation
0x1a    BYTE     Retrieve single byte from word

20s: SHA3

0x20    SHA3    Compute Keccak-256 hash

30s: Environmental Information

0x30    ADDRESS       Get address of currently executing account
0x31    BALANCE       Get balance of the given account
0x32    ORIGIN        Get execution origination address
0x33    CALLER        Get caller address. This is the address of the account that is directly responsible for this execution
0x34    CALLVALUE     Get deposited value by the instruction/transaction responsible for this execution
0x35    CALLDATALOAD  Get input data of current environment
0x36    CALLDATASIZE  Get size of input data in current environment
0x37    CALLDATACOPY  Copy input data in current environment to memory
0x38    CODESIZE      Get size of code running in current environment
0x39    CODECOPY      Copy code running in current environment to memory
0x3a    GASPRICE      Get price of gas in current environment
0x3b    EXTCODESIZE   Get size of an account's code
0x3c    EXTCODECOPY   Copy an account's code to memory

40s: Block Information

0x40    BLOCKHASH   Get the hash of one of the 256 most recent complete blocks
0x41    COINBASE    Get the block's beneficiary address
0x42    TIMESTAMP   Get the block's timestamp
0x43    NUMBER      Get the block's number
0x44    DIFFICULTY  Get the block's difficulty
0x45    GASLIMIT    Get the block's gas limit

*Figure 1 EVM INSTRUCTION SET – PART 1*

powerful ledger primitives and secure interaction with the outside world. This facilitates creation of

50s Stack, Memory, Storage and Flow Operations

```
0x50   POP       Remove item from stack
0x51   MLOAD     Load word from memory
0x52   MSTORE    Save word to memory
0x53   MSTORE8   Save byte to memory
0x54   SLOAD     Load word from storage
0x55   SSTORE    Save word to storage
0x56   JUMP      Alter the program counter
0x57   JUMPI     Conditionally alter the program counter
0x58   PC        Get the value of the program counter prior to the increment
0x59   MSIZE     Get the size of active memory in bytes
0x5a   GAS       Get the amount of available gas, including the corresponding reduction
0x5b   JUMPDEST  Mark a valid destination for jumps
```

60s & 70s: Push Operations

```
0x60   PUSH1   Place 1 byte item on stack
0x61   PUSH2   Place 2-byte item on stack
...
0x7f   PUSH32  Place 32-byte (full word) item on stack
```

80s: Duplication Operations

```
0x80   DUP1   Duplicate 1st stack item
0x81   DUP2   Duplicate 2nd stack item
...
0x8f   DUP16  Duplicate 16th stack item
```

90s: Exchange Operations

```
0x90   SWAP1   Exchange 1st and 2nd stack items
0x91   SWAP2   Exchange 1st and 3rd stack items
... ...
0x9f   SWAP16  Exchange 1st and 17th stack items
```

a0s: Logging Operations

```
0xa0   LOG0   Append log record with no topics
0xa1   LOG1   Append log record with one topic
... ...
0xa4   LOG4   Append log record with four topics
```

f0s: System operations

```
0xf0   CREATE       Create a new account with associated code
0xf1   CALL         Message-call into an account
0xf2   CALLCODE     Message-call into this account with alternative account's code
0xf3   RETURN       Halt execution returning output data
0xf4   DELEGATECALL Message-call into this account with an alternative account's code, but persisting the current values for `sender` and `value`
```

Halt Execution, Mark for deletion

```
0xff   SELFDESTRUCT   Halt execution and register account for later deletion
```

*Figure 2` EVM INSTRUCTION SET – PART 2*

ledger based smart contracts and provides a secure interface to the outside world.

The core business logic beyond ledger and block chain primitives and data communications with the outside world is then realised with standard stack machine instructions. Because this standard stack machine instruction set is Turing complete any kind of algorithm and data structure can be implemented from first principles. This provides great flexibility and allows higher level languages to be ported to the EVM using special purpose compilers that generate machine code for the EVM. Because the EVM is simple, standard stack machine experience and patterns for code generation for microprocessors can be reused when porting higher level languages to the EVM.

One such Higher level language is Solidity. The documentation for Solidity [SOLIDITY01] describes it as "a contract-oriented, high-level language whose syntax is similar to that of JavaScript and it is designed to target the Ethereum Virtual Machine (EVM)." Being, "statically typed," it " supports inheritance, libraries and complex user-defined types among other features." Using Solidity, the documentation continues, "it is possible to create contracts for voting, crowdfunding, blind auctions, multi-signature wallets and more". Figure 3 shows a very

```solidity
pragma solidity ^0.4.0;

contract SimpleStorage {
    uint storedData;

    function set(uint x) {
        storedData = x;
    }

    function get() constant returns (uint) {
        return storedData;
    }
}
```

Figure 3 - A Simple Example Contract in Solidity

simple example contract written in Solidity. This allows participants to set the value of a

Using the EVM designers and implementers of smart contracts have complete freedom how to structure, represent and encode the data to be used in the context of such a contract.

## 2.2 A Semantic Blockchain Procedural Language and Database – The HyperLedger Approach

In contrast to the EVM, HyperLedger does not provide stack machine or other low level virtual machine but instead provides a Byzantine Fault Tolerant ledger machine based on Liskov and Castros [Liskov et al 1999] Practical Byzantine Fault Tolerance (PBFT) algorithm. This machine can be accessed by external programs via a Web API making it easy to create complex real-life solutions with embedded smart contracts and secure distributed ledgers.

Figure 4 - Simple Chain Code for initializing a Ledger

The actual logic for smart contracts or other block chain ledger based functionality is implemented in what are called "Chain Code" modules. Chain code modules can be written in GO, a modern imperative language suitable for robust high-performance systems applications. Chain code modules consist of standard GO code but utilize a small API that exposes.

The example chain code program in Figure 3 illustrates how Chain code combines GO and the HyperLegder API. There are two key blocks of statements in the Example in Figure 3. The first block is

```
account = args[0]
accountValue, err = strconv.Atoi(args[1])
```

This takes the name of the account to be updated from the first argument ( `args[0]` )of the call invoking this chain code procedure and stores it in the variable `account`. It then takes the initial balance for the account from the second argument ( `args[1]` )of the call and stores it in the variable `accountValue`.

The second key block is then using the chain code API

```
// Write the state to the ledger
err = stub.PutState(account, []byte(strconv.Itoa(accountValue)))
```

```go
func (t *CrowdFundChaincode) Init(stub shim.ChaincodeStubInterface, function string, args []string) ([]byte, error) {
    // State variable "account"
    var account string
    // The value stored inside the state variable "account"
    var accountValue int
    // Any error to be reported back to the client
    var err error

    if len(args) != 2 {
        return nil, errors.New("Incorrect number of arguments. Expecting 2.")
    }

    // Initialize the state variable name
    account = args[0]
    // Initialize the state variable value
    accountValue, err = strconv.Atoi(args[1])
    if err != nil {
        return nil, errors.New("Expecting integer value for account initialization.")
    }

    fmt.Printf("accountValue = %d\n", accountValue)

    // Write the state to the ledger
    err = stub.PutState(account, []byte(strconv.Itoa(accountValue)))
    if err != nil {
        return nil, err
    }

    return nil, nil
}
```

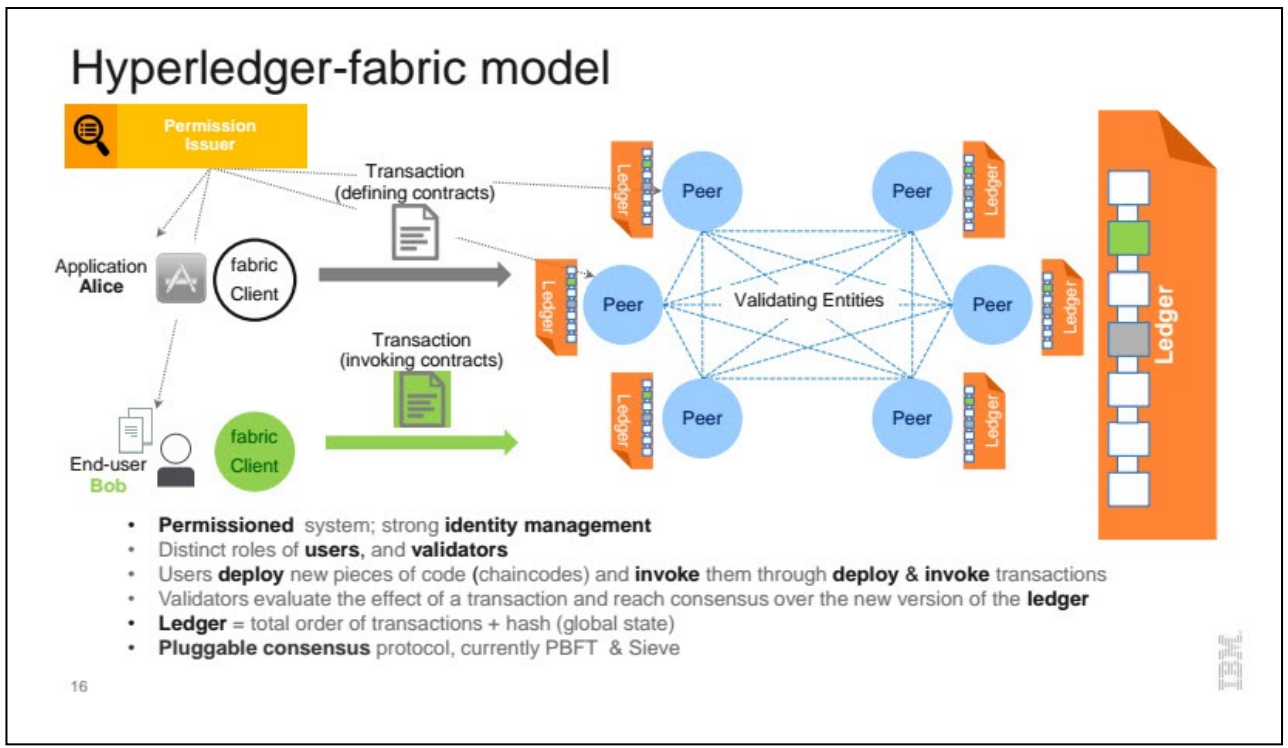

*Figure 5 - HyperLedger Structure from [BlockGeeks01]*

## 2.3 Problems with Procedural Semantics

Programs with procedural semantics make step by step (instruction by instruction) changes to variables representing the state of a program, machine or contract. This is true for all such programs whether blockchain related or otherwise. While such programs can feel intuitive to the creator, understanding them requires a process that synthesises meaning by comparing parts of a process with matching process fragments for which a meaning is already known. Synthesizing meaning in this way in the general case is a hard problem for both humans and machines. When Alan Turing sketched out the TURING MACHINE in his paper [TURING 1936] on the computability of numbers he created the basic underlying semantics for all subsequent procedural languages. Turing's paper used the device of the Turing Machine to prove that there is no general algorithm for determining if an algorithm ever finishes. While this is not the same as comprehending or understanding an algorithm or Programs with procedural semantics it is certainly one important aspect. Procedural programs meet to conform to very strong and restrictive assumptions in order to even just easily verify that they complete in a certain time for any possible input. Any algorithm for synthesizing the meaning of a program is neither not guaranteed to work for all possible algorithms nor in the general case guaranteed to ever finish even if the meaning is synthesisable.

This is important because it means that it is not in general possible to compare two smart contracts written as procedural code, to see if they have the same meaning unless they are essentially carbon copies of each other. It also means that it is not in general possible to automatically verify if a smart contract realised in procedural code meets certain specifications nor is it possible to guarantee such a check can be conducted in within some period.

This means that even if a procedural smart contract uses a carefully selected ontology for representing all information used by the smart contract throughout its life such smart contracts and their meaning cannot in general be compared, verified or understood automatically. In small scale or tightly locked down applications where e.g. all smart contract types are known upfront this is not a big problem since it is possible to select only smart contracts for which their meaning and other characteristics can be easily enough synthesised or computed even if this involves some considerable human intervention and art is some cases. This semantic meta data can then be tied to the smart contract like a manifest and used when reasoning over one or more such smart contracts.

In a more open environment where new smart contract types can be created by participants any time this is still of some help but requires a significant level of resources and sophistication and arguably some centralised governance to ensure compliance with minimum standards for semantic metadata manifests that are required for each smart contract in this case. This limits the usefulness of the semantic meta data manifest work around to special types of open environments where these factors are present.  Semantic block chains mechanisms with declarative semantics, as will be shown in the next section are designed to overcome this issue and ensure that every aspect of a smart contract has a clear meaning that can be analysed and understood by machines.

## 3  Semantic Blockchain with Declarative Semantics

Declarative languages sometimes have the reputation of being both difficult and offering lower performance. It is important to highlight here that while there are declarative languages that are either say like say ERLANG, APL or even both like say PROLOG there are clear examples that prove this does not need to be the case. The query language SQL for instance is both easy to use and in many cases also delivering very high performance.

Declarative Languages come in a much greater variety than procedural languages. They include languages based on the evaluation of functions such as LISP and its descendants, logic languages like PROLOG, Concurrent Guarded Horn Clauses and ERLANG, query languages like SQL as well as countless domain specific languages.  See also **[WIKIPEDIA03].** Two other classes of declarative formalisms for computation are actor based programs and state machine or state chart based programs.

The big benefit of declarative languages is that in many cases they are designed to make it much easier to reason about the meaning of a given program because they directly represent an ontology of the desired results. This is the case because declarative programs are representations of the required results in contrast to procedural programs that are representations of the steps intended to make changes to variables that will once all complete deliver a result.  If semantic block chain machines like Smart Contracts were encoded in a declarative language it would be much easier to compare and validate and even understand them in a more general way.  It is therefore worth considering what alternatives exist to realise block chain programmes with declarative semantics.

### 3.1 Alternatives for Declarative and Concurrent Declarative Computing

Functional Languages like LISP and its descendants are very well understood, easy to moderately difficult to use and reason about because everything you need to know tends to be in the same textual context. They are designed to make it relatively easy to create programmes where the semantics of the desired solution is directly readably because the program can operate directly on ontological statements and instance. Data and messages are in represented by the same formalism as the active code, functions that operate on data and messages.

Logic Languages like PROLOG are also well understood but often somewhat harder to use because you often need to understand context that is more widely dispersed and its formalism is more abstract. Good logic programmes though can also be excellent literary representations of the meaning in a similar way to functional language programmes. Again, similarly logic programs use the same mechanism for representing data and messages.

The big difference however is in the way  functional and logic languages are executed.  Functional languages can be compiled to be executed almost at machine level while logic languages require unification which does not lend itself to compilation at such a low level.

When it comes to supporting concurrency both functional and logic languages have been successfully extended or transposed to support the creation of programs with high levels of concurrency.  SEE GHC.

Concurrent functional programmes are very hard to reason about. They are more robust that concurrent procedural programs but require substantial experience and skill to understand.

Concurrent logic programmes like GHC's tend to be easier since there is a clear separation between concurrency and the non-concurrent parts.

Two much better formalism however are available to deal with concurrency in a declarative way. Actor based programming and State machine based Programming. In actor based programs most logic is encapsulated in small well-defined actors, who can only communicate via messages. Communication sequential processes HOARE GO are also very similar to this but emphasise communication pipes while actors have inbuilt individual and or shared mailboxes. Actors lend themselves define the action semantics in a very declarative way. Being message based the entire action protocol is directly formed of semantic building blocks from the domain concerned. Only the messages handlers usually revert to procedural logic.

State machines can represent even more of the action semantics in a declarative form. In addition to data and protocol they also represent state transition in a declarative way. Only when it comes to processing messages etc do the common examples like QUANTUM require the use of procedural logic. Since none of the existing formalism lends itself directly to a fully declarative approach in a scenario were concurrency integral to almost every problem its is worth exploring if a blending of approaches would yield a better solution.

## 3.2 Closing the Semantics Gap in Blockchain Computation

Both Functional and logic languages can serve as host or be injected into state machines and actors. Actors can also in themselves be easily represented using state machines. A combination of actors and state machines also enriches the state machine model with a clear formalism for which actor/state machine talks to which other actor/state machine.

When using a functional language paradigm e.g. a LISP like language everything can be easily encoded in this standard form. Data, messages, declarative functions, where appropriates also small less declarative functions snippets, every part of the state machine definition f for every actor rom states to declarative transition functions as well as all the additional aspects of actors beyond the embedded state machine.

A final interesting observation is that this combined formalism lends itself very well for the implementation of state machine replication such as the implementation of Practical Byzantine fault tolerance (PBFT) and newer protocols that implement Byzantine fault tolerant state machine replication with different characteristics.

Earlier in the chapter it has been shown that each of the different Blockchain mechanisms provides a good solution to the need for Certainty of Agreement. Using carefully constructed ontologies for all data also goes a long way towards meeting the need for Certainty of Meaning but still leaving an important gap when it comes to the meaning of computed results or interactions. The observations in the previous paragraph raises the question if this gap could be closed using a combination of declarative formalisms. The next section will show one approach that demonstrates how this gap can be closed.

## 3.3 Putting it all together - The Huuzlee Approach

Creating end to end declarative semantics for block chain applications was a key aim of the initial Huuzlee research project. With end to end declarative semantics block chain applications like smart contracts running on the Huuzlee platform could be more easily created, verified and compared. This open up a whole new space of applications in situations that require a very high degree of safety, security and robustness with complex chains of transactions / smart contracts, clear permissions for every capability or action but no single central authority that can enforce a monolithic solution. With end to end declarative semantics participants can assess applications from other participants and enforce clearly defined policies before letting such an application participate in a trading chain. Creation of new smart contracts also becomes easier and safer as their definition can be based on a well defined domain ontology and meaning and characteristics more easily validated using both expert reviews and automated tools.

A key observation that led to the chosen architecture was that real-life transaction chain almost always require multiple "actors" to co-operate. Often actors in a particular position in a chain can and often even need to have different but interchangeable implementations with respect to the chain but each implementation addressing different local circumstances. As a consequence, the Huuzlee project chose the actor concept as its fundamental atomic unit. Each actor implements an agreed protocol that is understood by other actors upstream and downstream in chains but completely hides its internal implementation or state otherwise. This makes actor implementations entirely interchangeable. With declarative definitions of actors it is not only easier to verify compliance with the set protocol but also to create a manifest of other characteristics that makes it easier to ensure that an appropriate implementation of an actor to comply with local requirements.

In order to define actors and their protocol in a declarative way it was necessary to find a formalism that would allow each state an actor can be in to be clearly defined in declarative terms using a suitable domain specific ontology. In addition, the condition for every transition of an actor from one state to another and all the resulting messages and changes to data should be equally defined in declarative terms using a suitable domain specific ontology. This made the well-known state machine formalism an obvious choice. State machines have a set of clearly named states with each state accepting one or more messages types. For each acceptable message in a given state there is a clear definition of the resulting change in terms of stored data and messages sent to other actors or even the actor concerned itself. In addition, each message received does lead to a clearly define transition either to the same state or another state. This atomic transaction like behaviour makes state machines more robust and easier to understand in this context than other possible approaches. It also explicitly exposes the protocol(s) a state machine or an actor defined using it adheres to.

Finally it was necessary to select a lower level formalism or declarative host language to use for data and message definitions and the declaration of conditions and the declarative definition of transformation function mapping messages received into changes in stored data and messages sent out in response.

Although a logic language similar to say PROLOG would have been a possible choice a functional language based on the well know LISP family of languages could be shown to be a better choice. Amongst the factors in its favour is the very natural way the language allows designers to define semantically well-defined message and data structures. In a similar way this type of language makes it equally easy to represent every part of a state machine specification in the same way as

```
ACTOR {

  DATA {
    buyoffer {
      product            { ? },    price       { ? },    quantity  { ? },
      buyer    { ? },              seller      { ? }                      },
    selloffer {
      product            { ? },    price       { ? },    quantity  { ? },
      buyer    { ? },              seller      { ? }                      },
    contract {
      product            { ? },    price       { ? },    quantity  { ? },
      buyer    { ? },              seller      { ? }                      }
  }

  MODEL {

    Initially {
            #Enter {
              transitionTo { $OPEN }
            },

            #Exit { 'do nothing' }
    },
    Open {
            #Enter { 'do nothing' }

            #buyoffermsg {
                    map { *THIS, @buyoffer },
                    match { @selloffer, @buyoffer,
                     @SUCCEEDS {
                            transitionTo => $Closed
                     },
                     @FAILS
                            transitionTo => _
                    }
              }
            },

            #selloffermsg {
                    map { *THIS, @selloffer },
                    match { @selloffer, @buyoffer,
                     @SUCCEEDS {
                            transitionTo => $Closed
                     },
                     @FAILS
                            transitionTo => _
                    }
              }
            },

            #Exit {
                    send { @contract.buyer,
                       compose >>> Contract Notice: Buy @contract.quantity unit
                              --> of @contract.Product at @contract.price
                              --> from @contract.seller <<<
                    }
                    send { @contract.seller,
                       compose >>> Contract Advice: Sell @contract.quantity unit
                              --> of @contract.Product at @contract.price
                              --> to @contract.buyer <<<
                    }
            }
    },

    Closed {
            #Enter { terminateActor }
    }
  }
}
```

*Figure 6 - Simple Smart Contract in Huuzlee*

directly in the host language or the adapted state machine formalism can be easily represented in the same way.

Figure 6 provides a simple example of a smart buy/sell contract. The contract is implemented as a Huuzlee actor. Once created it waits for a matching pair of buy and sell messages from a single buyer and seller pair. Buyer and seller can revise their offer as often as they like until they have both sent a matching offer at the same time. In a real life situation many other features may be present such as alerting the other side of the latest offer/counter offer etc but as a basic skeleton this simple example allows us to create a fully functional contract in a declarative way. The names for data elements, states etc have not been explicitly cross referenced to a formal ontology to keep the example short but again in real life this would be done and would make every aspect of this contract semantically fully specified.

It is worth noting that the smart contract in the example does not in itself reference an underlying block chain mechanism in any way. It would for instance be absolutely fine to test the basic logic of the contract using a single execution node – completely omitting byzantine fault tolerance and other block chain aspects under those circumstances. This makes it much easier to test more complex contracts and protocol chains with minimum overhead first. Once ready this same contract without any changes whatsoever can then be deployed to an arbitrary complex execution note network that replicates it on all active nodes and executes every message receipt in ion lockstep on all nodes with encrypted voting to ensure the integrity of every replica thereby achieving full byzantine fault tolerance.

While the ability to test in a non replicated instance is very useful this approach has a much more fundamental impact. By separating the execution mode from the smart contract itself it is possible to develop a contract once and then deploy it for instance into different regulatory or contractual situation requiting different sets of replication nodes.

In a cross border payments scenario for instance we may find the following *requirements:*

Any sovereign nation in this example can supervise any participants and transactions that fall under its sovereignty to ensure that all applicable laws can be enforced and be certain that the ability of any participant under its own sovereignty to conduct business with participants from another nation that also permits such business cannot be interfered with by a third nation or other actor. End-to-End transaction integrity and privacy will be assured by design at a network level even under adverse conditions such as where one or more nodes have come under the control of a malicious attacker.

To achieve this the smart contracts needed for the payments chain would run on byzantine fault tolerant node networks. Policies for a transaction applied by a direct participant in a transaction such as a sender, recipient or any service provider in the chain and indirect participants such as e.g. regulators will be enforced automatically and intrinsically as the transaction is executed.

A permanent record of all attempted, completed, failed and live transactions is retained by the blockchain – keeping a permanent record of all completed and failed smart contract instances.

The architecture meets those requirements through a design using a permission based semantic block chain created within an open network of attack resistant self-healing smart execution nodes capable of executing smart contracts.

Smart contracts in this network are a composite of the sender's and receiver's transaction specification overlaid with self-enforcing policies put in place by service providers and regulators throughout the end to end transaction chain.  In a typical real-world scenario, the architecture would ensure that a minimal set of nodes for a direct remittance from one country to another would include at least one node for the service provider in the sender country and the same for the provider in the recipient country plus at least additional ledger node in both sender and receiver.

This scenario can be readily implemented with the Huuzlee approach because smart contracts can be assembled from small behaviour components such as country specific policies. Behaviours can be reused across different smart contract if the same rules apply. Participant specific behaviours can be mixed in to treat e.g. whole sale participants with a different rule set from that used for retail participants.  In a high volume cross border scenario it is easier for affected regulators to asses or audit smart contract because their executable specifications can be directly analysed and application of specific rules applicable can be verified directly.

## 4   Conclusion:  Comparing Blockchain Procedural and Declarative Semantics

It is worth now to compare Block Chain mechanisms with those that have declarative semantics side by side.  This is done in the following table:

| Characteristic | Blockchain with Procedural Semantics | Blockchain with full Declarative Semantics |
|---|---|---|
| **Examples** | 1. Stack machine based formalisms like Ethereum<br><br>2. BFT based formalism like HyperLedger | 1. State Machine based formalisms with ontology based data definitions like e.g.  Huuzlee |
| Semantics of smart contracts implemented in the formalism can be readily verified against formal specifications using automatic tools. | **No.**<br><br>The behaviour is implicit, and specifications can be reverse engineered only in very limited cases | **Yes.**<br><br>The behaviour and meaning is explicitly specified and can be compared to specifications in a suitable declarative format. |
|  |  |  |
| The formalism can be used to specify the meaning and behaviour of a class of smart contract. | **No.**  The formalism does not allow the explicit specification of meaning and behaviour. | **Yes.**  The formalism is designed to explicitly specify behaviour and meaning. |
| The formalism makes it easy to prevent unintended side-effects or race conditions in the implementation of smart contracts. | **No.**  Implicit behaviour and complex state representations in sets of values makes this very difficult. Mutable variables exacerbate this problem | **Yes.**  Crisp- human readable state model transition models, transaction style all or nothing modification of state and the lack of mutable variables eliminate many unintended side effects and race conditions and make it easier to identify any remaining. |
| The formalism makes it easy to compose suitable building block components into complete implementations of smart contracts. | **No.**  Implicit behaviour and complex state representations (especially with mutable variables) make composition from partial building block components unsafe in the general case. | **Yes.**  Hierarchical state models make behaviour composition easy, readily verifiable and well suited in safety critical conditions. Explicit definition of the meaning of values and the ability to explicitly compose structures of values also simply the safe composition of values and the automatic verification of compositions. |
| How scalable are services based on smart contracts implemented in the formalism? | **Moderately Scalable.**  Scalability is limited at a transaction level because of the restricted composability of transaction behaviour components. | **Highly Scalable.**  Scalable at all levels because transaction can be composed freely thus scaling easily even to very complex individual transactions. At a service level individual services can be federated thus making services safely scalable even at high volumes. |
| | | |
| Does the formalism allow an evolution of the block chain protocol to easily address future security challenges ? | **No.** The block chain implementation is hard coded. | **Yes.** The block chain implementation is using the formalism itself and can be transparent to the next layer of functionality thus allowing an evolution of the protocol used or even several slightly different implementations of the same protocol at the s |
| Does the formalism allow different implementation of a given block chain protocol to be used side by side to make it harder for an attack to succeed? | **No.** The block chain implementation is hard coded. | **Yes.** The block chain implementation is using the formalism itself thus allowing several slightly different implementations of the same protocol to be used side by side |
| Does the formalism allow the use of arbitrary ontologies as an integral part of the definition of a smart contract? | **No. Some Workarounds** The block chain implementation does not recognise ontologies as a first class object thus – no. It is however possible to create a work around for this by creating alternative procedures, testing token/signal type and then selecting the appropriate procedure to apply. | **Yes. Native support** Ontologies are a first class object of the formalism and any value is clearly identified by an ontology. In addition, every behaviour (smart contract component / state machine) is also clearly identified by a behaviour. This means that every semantic aspect of a smart contract can be clearly semantically assed and even readily verified by automatic means. |
| How new is the formalism and how widely is it used. | **Have been available for some time – widely used.** The formalism in this category have been in use for some time and consequently are widely used. This means they can be readily employed for applications for which their | **New – No yet widely used** The formalism in this category are relatively new and not yet widely used. Before using this approach in a production scenario, it is important to plan for an extended period to |

| | comparative disadvantages do not matter. | validate any specific implementation. |
|---|---|---|

It is clear from the table above that while semantic block chain designs with procedural semantics can very suitable for domains with limited semantic and compositional complexity they suffer from significant challenges in domains that go beyond these limits.  Block chain design with declarative semantics on the other hand avoid those challenges but are relatively new.  It is worth noting though that the component approaches used in Block chain design with declarative semantics are widely used in safety critical applications and thus at that level much knowledge and experience exists.  Nevertheless, further research would benefit block chain design with declarative semantics.

http://blockchain.cs.ucl.ac.uk/barac-project/

https://www.researchgate.net/publication/317492448_CHANGING_BANKING_LANDSCAPE_THE_PSD2

Complexity of mapping all semantic content; manual, needs domain knowledge, constantly changing and needs agreement between different parties within and across organizations – need to use cognitive computing and needs to be self sustaining -> HOW TO ADDRESS THIS?